\documentclass[preprint,12pt]{STNI}
\date{\today}

\usepackage{graphics}
\usepackage{amsmath}

\biboptions{square,sort&compress}

\begin{document}

\title{A simple formula for local burnup and isotope distributions based on approximately constant relative reaction rate}

\author[rvt]{Cenxi Yuan\corref{cor1}}
\author[rvt]{Xuming Wang}
\author[rvt]{Shengli Chen}

\address[rvt]{Sino-French Institute
of Nuclear Engineering and Technology, Sun Yat-Sen University,
Zhuhai, 519082, Guangdong, China}

\cortext[cor1]{yuancx@mail.sysu.edu.cn}

\begin{abstract}
A simple and analytical formula is suggested to solve the problems of the local burnup and the isotope distributions.
The present method considers two extreme conditions of neutrons penetrating the fuel rod. Based on these considerations, the formula is obtained to calculate the reaction rates of $^{235}$U, $^{238}$U, and $^{239}$Pu and straightforward the local burnup and the isotope distributions. Starting from an initial burnup level, the parameters of the formula are fitted to the reaction rates given by a Monte Carlo (MC) calculation. Then the present formula independently gives very similar results as the MC calculation from the starting to high burnup level, but takes just a few minutes. The relative reaction rates are found to be almost independent on the radius (except $(n,\gamma)$ of $^{238}$U) and the burnup, providing a solid background for the present formula. A more realistic examination is also performed when the fuel rods locate in an assembly. A combination of the present formula and the MC calculation is expected to have a nice balance on the accuracy and the cost on time.
\end{abstract}

\maketitle

\section{\label{sec:level1}Introduction}
To increase the efficiency of the fuel, one possible way is to increase the burnup of the fuel before discharge. The local burnup on the edge of the UO$_{2}$ fuel rod is much higher than the average burnup. Thus it is of great importance to investigate the properties of the rim of the fuel rod when considering the increment of the average burnup. Many investigations show that the mechanical structure close to the surface is rather different from that at the center of the fuel rod. In the high burnup range, a microstructure change was found on the rim of the fuel rod through transmission electron microscopy~\cite{Nogita1995}. One explanation of the formation of the high burnup structure supposed that the bubbles in high burnup region are nucleated and stabilized by fission fragments, which depends on the fission rate~\cite{Lee2000}. The small pores at the high burnup region are calculated to be highly overpressurized~\cite{Koo2001}. Recently, the UO$_{2}$ fuel in both light and heavy water reactor are investigated to understand the high burnup structure of the fuel~\cite{Noirot2008,Amaya2009}.

In a thermal reactor, the neutrons generated from fission need to be slowed down in the moderator to be able to induce next fission. When the low energy neutrons go from the moderator to the fuel rod, they are firstly absorbed by the fuel close to the surface. In general, the reaction rate is higher in the rim when induced by slowed neutrons, such as the $(n,f)$ reaction of $^{235}$U and $^{239}$Pu induced mainly by the thermal neutron. In contrast, the (n,$\gamma$) reaction of $^{238}$U is mainly induced by the resonance neutrons. The local burnup phenomena is mainly caused by the $(n,\gamma)$ reaction of $^{238}$U, which has large cross section for resonance neutrons, especially some high peaks of cross section at certain energies. Such reaction produces much more $^{239}$Pu near the surface of the fuel rod, thus giving origin to the increased local burnup. As a consequence of the different reaction rates along the radial direction, the U and Pu isotopes also have different radial distributions. In a fast reactor, the Pu isotopes are redistributed along the radial direction~\cite{Marcello2012}. A large local burnup is not expected because the neutrons do not need to be slowed down in the moderator.

In nuclear science, theoretical models are very important and powerful to solve various of problems. Our previous works have shown that the mass, levels, electromagnetic moments and transitions, and other properties of the nuclei can be described with high precision in the frame of the nuclear shell model even for very unstable nuclei, such as $^{22}$C and $^{21}$Al~\cite{yuan2012}.
In nuclear reactor, the simulation is even more important because the experimental data are relatively difficult to be obtained, sometimes with high risk. For the present problem, various models can be used to calculate the local burn up and the isotope distributions along the radial direction. The TRANSURANUS model is very successful in the description of the local burnup and related properties in various reactors including light and heavy water reactors~\cite{Lassmann1994,Lassmann1998,Schubert2008}. In TRANSURANUS model, a fixed neutron flux is assumed except for the absorption reaction of $^{238}$U and $^{240}$Pu~\cite{Schubert2008}, while the RAPID model considers detailed properties of the neutron flux at each burnup~\cite{Lee20002}. Recently, an empirical formulation is suggested to describe the local burnup and high burnup properties based on DIONISIO code~\cite{Soba2013,Soba2014,Lemes2015}.

Here we suggest an analytical and simple formula to calculate local burnup and isotope distributions based on the constant relative reaction rates at the different radius (except for $^{238}$U) and burnup. The present method considers the radial distribution of the neutron flux through two extreme conditions. The parameters are fitted to the reaction rates of the starting burnup level. From this level, the present method are shown to have independent and similar results as the Monte Carlo (MC) simulation, but just a few minutes is needed because its simplicity.

\section{\label{sec:level2}The description of the model}
The rate of certain reaction type ($x$) between neutrons and a given nuclide ($A$) atan average burnup ($bu$) can be written as a function of the radius ($r$) and the energy of neutron ($E$).
\begin{eqnarray}\label{RR1}
 R(r,A,x) &=& \int_{E} N(r,A)\phi(r,E)\sigma(A,x,E)dE,
\end{eqnarray}
where $N(r,A)$ is the concentration of the nuclide $A$ at the radial position $r$, $\phi(r,E)$ is the neutron flux at the energy $E$ and the radius $r$, $\sigma(A,x,E)$ is the microscopic cross section between neutron and $A$ for reaction type $x$ at the energy $E$. The $(n,f)$ and $(n,\gamma)$ reactions are assumed to be the two more important reaction types in the fuel rod. In the present work, the main purpose is to examine the validity of the present model. Thus only the $^{235}$U, $^{238}$U, and $^{239}$Pu isotopes are considered to constrain the model testing to a simple case.

In the present study, the Eq.~(\ref{RR1}) is assumed to be two terms for $(n,f)$ and $(n,\gamma)$ of $^{235}$U and $^{239}$Pu, and the $(n,f)$ of $^{238}$U:
\begin{eqnarray}\label{RR2}
 R(r) &=& N(r)[c_{1}+c_{2}e^{b_{1}(r-r_{0})}],
\end{eqnarray}
and three terms for $(n,\gamma)$ of $^{238}$U:
\begin{eqnarray}\label{RR3}
  R(r) &=& N(r)[c_{3}+c_{4}e^{b_{1}(r-r_{0})}+c_{5}e^{b_{2}(r-r_{0})}],
\end{eqnarray}
where the coefficients $c$ are specified for each nuclide and the reaction type. The coefficient $b_{1}$ is expected to be the same for all three nuclides. The coefficient $b_{2}$ in the $(n,\gamma)$ of $^{238}$U is much larger than $b_{1}$.

The following discussion concentrates on the explanations of the above assumptions of the reaction rates. In a thermal reactor, two kinds of the neutron reactions are most important, the scattering reaction in the moderator and the absorption reaction in the fuel rod. When the slowed down neutrons penetrate into the fuel rod, we mainly concentrate on the neutron flux at the thermal and resonance region which are important to the absorption reaction in the fuel rod. The average energy of fission neutron is around $2$ MeV, which is much larger than that of the thermal and resonance neutron. Thus no multiplication of the neutron flux is considered at the energy of thermal and resonance region in the following discussion. Two extreme conditions can be considered for certain energy of the neutron. The first one is that the neutrons are strongly absorbed by the fuel, which indicates that the mean free path (MFP) of the absorption, $\lambda(a,E)=\frac{1}{\Sigma_{a}(E)}=\frac{1}{\Sigma_{a,5}(E)+\Sigma_{a,8}(E)+\Sigma_{a,9}(E)}$, is much smaller than the size of the fuel rod at the energy $E$. The second one is that few neutrons are absorbed in the fuel at the energy $E$.

For the first extreme condition, one can further simplify that the velocity of the neutrons are all perpendicular to the surface of the fuel rod. Such assumption is acceptable because the velocity is almost symmetric around the radial directions in the real case. Then the transportation of the neutrons at certain energy $E$ is the solution of the Boltzmann equation in the cylindrical coordinate without the terms of the scattering and the source. The radial part of the equation is:
\begin{eqnarray}\label{Bo-Eq-1}
 -\frac{1}{r}\frac{\partial r\phi(r,E)}{\partial r}+ \Sigma_{a}(E)\phi(r,E) &=& 0,
\end{eqnarray}
where the negative sign comes from the opposite direction between the velocity and the radial vector, and the solution is:
\begin{eqnarray}\label{flux1}
 \phi(r,E) &=& \frac{r_{0}}{r}\phi(r_{0},E)e^{(r-r_{0})/\lambda(E)},
\end{eqnarray}
where $r_{0}$ is the radius of the fuel rod. Because of the assumption of the strong absorption, the neutron flux is meaningful near the surface of the fuel rod. The divergence at $r=0$ should not be considered. Because $\lambda(a,E)$ is much smaller than the size of the fuel rod, $r_{0}$, the neutron flux decreases very quickly to zero at small $(r-r_{0})$. The neutron flux $\phi(r,E)$ is approximately $\phi(r_{0},E)e^{(r-r_{0})/\lambda(E)}$ near the surface and zero at the other radial region. The reaction rate per nuclei $R(r,A,x,E)/N(r,A)$ is proportional to $e^{(r-r_{0})/\lambda(a,E)}$, resulting the third term in Eq.(\ref{RR3}). One example of such situation corresponds to the $(n,\gamma)$ reaction of $^{238}$U at certain energy. The magnitude of the atomic concentration of $^{238}$U is around $10^{22}$/cm$^{3}$ in the fuel rod. For $(n,\gamma)$ reaction of $^{238}$U, some peaks of cross section at the resonance region can achieve $10^{4}$ barn~\cite{nndc}, resulting $\lambda(a,E)=\frac{1}{\sigma_{a,8}(E)N_{8}}$ at the magnitude of $10^{-2}$cm, which is much smaller than the size of the fuel rod.

For the second extreme condition, the $\lambda(a,E)$ is much larger than the size of the fuel rod $r_{0}$. The $\phi(r,E)$ is almost unchanged in the fuel rod and $R(r,A,x,E)/N(r,A)$ is also approximately constant, resulting the first term in Eq.(\ref{RR2}) and (\ref{RR3}). The magnitude of the concentrations of $^{235}$U and $^{235}$Pu are around or less than $10^{20}$/cm$^{3}$ in the fuel rod. If the $\sigma_{a,5;a,9}(E)$ and the $\sigma_{a,8}(E)$ are much smaller than $10^{4}$ barn and $10^{2}$ barn, respectively, the $\lambda(a,E)$ is much larger than $r_{0}$.

For the situations between these two limits, the second term in Eq.(\ref{RR2}) and (\ref{RR3}) are assumed with $b_{1}$ the same magnitude of $\frac{1}{r_{0}}$. Please note that the above discussion is restricted in the fuel rod. If the neutrons go out of the fuel rod, they may be slowed and re-enter the fuel rod with a different energy.

In principle, $c, b$ in Eq.(\ref{RR2}) and (\ref{RR3}) can be calculated through the cross section data and the neutron flux at $r_{0}$. But in some energy regions, the cross sections changes dramatically, and hence it is difficult to obtain the coefficients. Spatially Dependent Dancoff Method can be used to calculated the cross-sections in the resonance region~\cite{Matsumoto2005}. In the present work, these coefficients are fitted using a MC calculation. An initial burnup level $bu_{1}$ is assumed and with it a MC calculation is done for one fuel rod to calculate the reaction rates at different radial positions. The coefficients are then fitted to these reaction rates and used to calculate the burnup levels after $bu_{1}$. After $bu_{1}$, the MC calculation and the present formula are performed independently for all burnup levels, $bu_{2}$, $bu_{3}$ and so on. The concentrations can be calculated through:
\begin{eqnarray}\label{RR4}
 N_{5}(r,bu_{i+1})&=& N_{5}(r,bu_{i})-(R_{5,f}(r)+R_{5,\gamma}(r))T, \nonumber \\
 N_{8}(r,bu_{i+1})&=& N_{8}(r,bu_{i})-(R_{8,f}(r)+R_{8,\gamma}(r))T, \nonumber \\
 N_{9}(r,bu_{i+1})&=& N_{9}(r,bu_{i})+(R_{8,\gamma}(r)-R_{9,f}(r))T,
\end{eqnarray}
by assuming that the reaction rates do not change during the time $T$. The time duration $T$ corresponds to the burnup change between two levels:
\begin{eqnarray}\label{deltabu}
\Delta bu&=& \frac{QT\int_{0}^{r_{0}} [R_{5,f}(r)+R_{8,f}(r)+R_{9,f}(r)]2\pi rdr}{M_{0U}(\pi r_{0}^{2})}. \nonumber
\end{eqnarray}
where $Q$ is the average energy released by fission and $M_{0U}(\pi r_{0}^{2})$ is the initial mass of U isotopes in volume $\pi r_{0}^{2}$.
One can transform:
\begin{eqnarray}\label{deltabuRR}
\Delta N_{5}(r)&=& \frac{M_{0U}(\pi r_{0}^{2})}{Q}\frac{-(R_{5,f}(r)+R_{5,\gamma}(r))}{\int_{0}^{r_{0}} [R_{5,f}(r)+R_{8,f}(r)+R_{9,f}(r)]2\pi rdr}\Delta bu. \nonumber \\
\Delta N_{8}(r)&=& \frac{M_{0U}(\pi r_{0}^{2})}{Q}\frac{-(R_{8,f}(r)+R_{8,\gamma}(r))}{\int_{0}^{r_{0}} [R_{5,f}(r)+R_{8,f}(r)+R_{9,f}(r)]2\pi rdr}\Delta bu. \nonumber \\
\Delta N_{9}(r)&=& \frac{M_{0U}(\pi r_{0}^{2})}{Q}\frac{(R_{8,\gamma}(r)-R_{9,f}(r))}{\int_{0}^{r_{0}} [R_{5,f}(r)+R_{8,f}(r)+R_{9,f}(r)]2\pi rdr}\Delta bu.
\end{eqnarray}
In MC calculation, the reaction rates in above equations are simulated with the concentration of all three isotopes at each burnup level. The present formula calculate all reaction rates through the coefficient fitted to the reaction rate at $bu_{1}$. The $R_{9,\gamma}(r)$ reaction rate is neglected in above equations  in both MC and analytical calculation because the present model does not include $^{240}$Pu. It is acceptable as present work concentrates on the examination of the formula not on a real burnup problem. The next section shows that the present formula can obtain results for $bu_{n}$ quite close to those obtained with the  MC calculation.

The above equation looks similar to the formula in TRANSURANUS model~\cite{Lassmann1994,Schubert2008}, in which the relationship between $N(r)$ and $bu$ can be generally written as:
\begin{eqnarray}\label{TRANSURANUS}
\frac{dN_{j}(r)}{dbu}&=& -\sigma_{a,j}N_{j}(r)f_{j}(r)A+\sigma_{c,j-1}N_{j-1}(r)f_{j-1}A,
\end{eqnarray}
where $j$ is one kind of nuclide. For $^{235,238}$U, there are no the second terms. For $^{236}$U, $^{237}$Np, and $^{238,239,240,241,242}$Pu, the $j-1$ are $^{235,236}$U, $^{237}$Np, $^{238}$U, and $^{239,240,241}$Pu, respectively. The $\sigma_{a}$ and $\sigma_{c}$ are the one-group effective cross sections for the total neutron absorption and neutron captured, respectively. The cross sections are obtain differently for UO$_{2}$ and MOX fuel because of the very different initial concentrations in each fuel and the corresponding different neutron spectrum. $A$ is a conversion constant. The $f(r)$ is the radial form factor, which is $1+p_{1}exp(-p_{2}(r_{0}-r)^{p3})$ for $^{238}$U and $^{240}$Pu and unit for all other nuclide. The $f(r)$ comes from the resonance absorption and the parameters are determined by comparison with measurements~\cite{Lassmann1994,Schubert2008}. The local burnup and isotope distributions can be calculated through Eq.~(\ref{TRANSURANUS}). More details can be found in Ref.~\cite{Lassmann1994,Schubert2008}. The present work considers not unit $f(r)$ (actually the different radial neutron flux) for all nuclei.

\section{\label{sec:level3}Calculations and discussions}

\begin{table}
\begin{center}
\caption{\label{start} Relative radius, atomic concentrations and relative reaction rates of each nucleus. The atomic concentrations are given in cm$^{-3}$, all the other quantities are dimensionless. }
\vspace{1mm} \setlength{\tabcolsep}{7pt}
\begin{tabular}{cccccccccc}
\hline
    $r'/r_{0}$  &  $N_{5}$  &   $N_{8}$ & $N_{9}$ & $R'_{5,f}$& $R'_{8,f}$& $R'_{9,f}$ & $R'_{5,\gamma}$& $R'_{8,\gamma}$   \\
\hline
1/4         & 6.65$\times 10^{20}$  &	2.17$\times 10^{22}$&	3.69$\times 10^{19}$  &   1.00  &    0.0025  &  2.50  &   0.23  &   0.0183  \\
5/8         & 6.63$\times 10^{20}$	&	2.17$\times 10^{22}$&	3.91$\times 10^{19}$  &   1.02  &    0.0024  &  2.49  &   0.23  &   0.0189  \\
13/16       & 6.61$\times 10^{20}$	&	2.17$\times 10^{22}$&	4.33$\times 10^{19}$  &   1.05  &    0.0024  &  2.49  &   0.23  &   0.0203  \\
29/32    	& 6.59$\times 10^{20}$	&	2.17$\times 10^{22}$&	5.01$\times 10^{19}$  &   1.07  &    0.0023  &  2.48  &   0.23  &   0.0230  \\
61/64    	& 6.58$\times 10^{20}$	&	2.17$\times 10^{22}$&	6.26$\times 10^{19}$  &   1.08  &    0.0023  &  2.47  &   0.23  &   0.0283  \\
125/128     & 6.58$\times 10^{20}$	&	2.16$\times 10^{22}$&	8.49$\times 10^{19}$  &   1.09  &    0.0022  &  2.47  &   0.23  &   0.0368  \\
127/128     & 6.57$\times 10^{20}$	&	2.16$\times 10^{22}$&	1.26$\times 10^{20}$  &   1.10  &    0.0022  &  2.47  &   0.23  &   0.0552  \\
\hline
\end{tabular}
\end{center}
\end{table}

In the present work, the continuous energy Monte Carlo code TRIPOLI-4~\cite{Petit2008} is used for the MC calculations as the starting point of the analytical formula and the reference for after calculations. The geometry of the fuel rod in the present investigation is set to be $r_{0}=0.4127$ cm, with cladding between  $0.4127$ and $0.4744$ cm. The moderator is in an outside box with the length $1.2647$ cm with all surface reflection all neutrons. The fuel rod is divided to seven parts in the MC simulation, with the dividing point located at $r/r_{0}=1/2, 3/4, 7/8, 15/16, 31/32, 63/64$. The corresponding center $r'/r_{0}$ of each part is listed in Table~\ref{start}. For a normal UO$_{2}$ fuel, there is no Pu isotopes at the beginning. It is reasonable to start present calculation at a certain burnup $bu_{1}$ with Pu included. The MC calculation is done with $3.3\%$ enrichment $^{235}$U fuel in the fuel cell to obtain $(n,f)$ and $(n,\gamma)$ reaction rates of $^{235}$U and $^{238}$U. With the reaction rates, the number of $^{235}$U, $^{238}$U, and $^{239}$Pu in a certain burnup can be calculated by assuming the reaction rates do not change in this period.
The concentration of $^{235}$U, $^{238}$U, and $^{239}$Pu at $bu_{1}=3.4$ MWd/kgU is given in Table~\ref{start} as the starting point of the following calculations.

The $(n,f)$ and $(n,\gamma)$ reaction rates of $^{235}$U, $^{238}$U, and $^{239}$Pu can be simulated by using concentrations listed in Table~\ref{start}.
The corresponding relative reaction rates defined below are also listed in Table~\ref{start}:
\begin{eqnarray}\label{RR5}
 R'_{5,f}(r) &=& \frac{R_{5,f}(r)/N_{5}(r)}{R_{5,f}(r=r_{0}/4)/N_{5}(r=r_{0}/4))} \nonumber \\
 R'_{8,f}(r) &=& \frac{R_{8,f}(r)/N_{8}(r)}{R_{5,f}(r)/N_{5}(r)}\nonumber \\
 R'_{9,f}(r) &=& \frac{R_{9,f}(r)/N_{9}(r)}{R_{5,f}(r)/N_{5}(r)} \nonumber \\
 R'_{5,\gamma}(r) &=& \frac{R_{5,\gamma}(r)}{R_{5,f}(r)} \nonumber \\
 R'_{8,\gamma}(r) &=& \frac{R_{8,\gamma}(r)/N_{8}(r)}{R_{5,f}(r)/N_{5}(r)} \nonumber
\end{eqnarray}
The above equations firstly scale the reaction rate by the corresponding concentration. Because the concentration is different at different radial position.  The scaling cancels such difference in the consideration of reaction rate. And then calculate the reaction rates relative to that of $(n,f)$ reaction of $^{235}$U. We find that the radial distribution of these relative reaction rates are almost constant except $(n,\gamma)$ reaction of $^{238}$U. Thus all the reaction rates except that of $(n,\gamma)$ of $^{238}$U can be treated in Eq~(\ref{RR2}) with the same coefficients. The radial distributions of these reaction rates are fitted as,
\begin{eqnarray}\label{RR6}
 R_{5,f}(r) &=& cN_{5}(r)f(r), \nonumber \\
 R_{8,f}(r) &=& 0.024cN_{8}(r)f(r), \nonumber \\
 R_{9,f}(r) &=& 2.49cN_{9}(r)f(r), \nonumber \\
 R_{5,\gamma}(r) &=& 0.23cN_{5}(r)f(r), \nonumber \\
 f(r) &=& 0.99+0.11e^{3.3(r/r_{0}-1)}, \nonumber \\
 R_{8,\gamma}(r) &=& 0.0183cN_{8}(r)[0.97+0.42e^{3.3(r/r_{0}-1)}+2.85e^{50(r/r_{0}-1)}],
\end{eqnarray}
where $c$ is the reaction rate per nuclei of $(n,f)$ of $^{235}$U at $r=r_{0}/4$ and canceled in Eq.~(\ref{deltabuRR}). Please note that the same $b_{1}$ is used for all nuclei. From $bu_{2}$, MC and analytical calculations are independent. It is straightforward to see the advantage of transforming the burnup problem to the reaction rate problem in the present work. The reaction rate actually is the most direct quantity to obtain the burnup and the concentration. In many models, the reaction rate must be obtained after solving the neutron flux problem. The present work treats the neutron flux problem in a simple way in section~\ref{sec:level2} and considers the relative reaction rate. The validity of these considerations will be discussed later.

\begin{figure}
\includegraphics[scale=0.50]{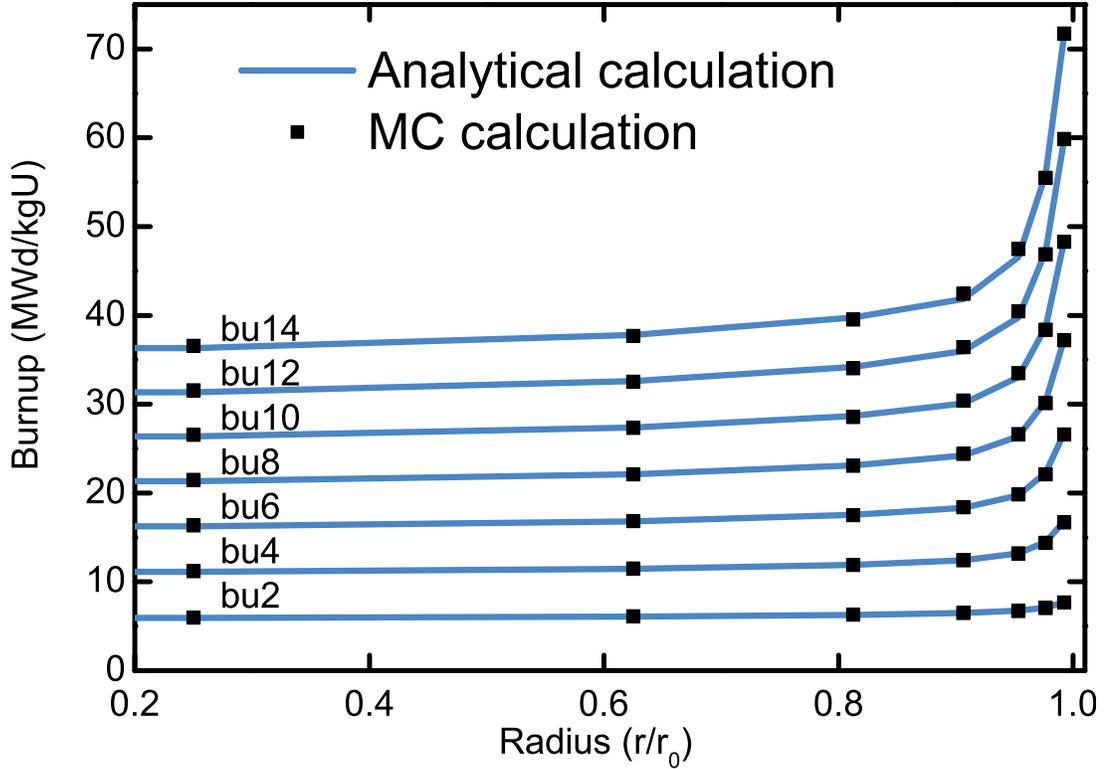}
\caption{\label{burnup} The comparison of the local burnup given by present formula and MC calculations.}
\end{figure}

\begin{table}
\begin{center}
\caption{\label{data1} The comparison of the local burnup (unit in MWd/kgU) at three levels given by present formula (AN) and MC calculations.}
\vspace{1mm} \setlength{\tabcolsep}{7pt}
\begin{tabular}{cccccccc}
\hline
    $r'/r_{0}$  &  $bu_{2,MC}$  &   $bu_{2,AN}$ & $bu_{8,MC}$ & $bu_{8,AN}$ & $bu_{14,MC}$ & $bu_{14,AN}$   \\
\hline
1/4         & 5.96    &	5.94    &	21.48 	&  21.33    &    36.57 &  	36.30    \\
5/8         & 6.11   	&	6.09    &	22.11 	&  22.09    &    37.69 &  	37.80    \\
13/16       & 6.29   	&	6.29    &	23.05 	&  23.11    &    39.53 &  	39.75    \\
29/32    	  & 6.48   	&	6.49    &	24.40 	&  24.22    &    42.42 &  	41.83    \\
61/64    	  & 6.72   	&	6.74    &	26.62 	&  26.37    &    47.48 &  	46.55    \\
125/128     & 7.06   	&	7.08    &	30.13 	&  30.38    &    55.50 &  	55.91    \\
127/128     & 7.68   	&	7.72    &	37.20 	&  37.31    &    71.75 &  	71.64    \\
average errors     &    	&	0.32\%    &	 	&  0.55\%      &     &  	0.84\%   \\
\hline
\end{tabular}
\end{center}
\end{table}

In the following discussion $\Delta bu$ is set to be $2.85$ MWd/kgU and the total burnup is calculated to $bu_{15}=43.30$ MWd/kgU. The local burnup from two calculations are presented in FIG.~\ref{burnup}. The analytical calculations can give almost the same results as MC calculations. The detailed data of three burnup leves in Table.~\ref{data1} shows that the average errors are less than $1\%$.

The radial distributions of U and Pu isotopes at three burunp levels are presented in Table~\ref{data2}. The $bu_{2}$ is the first burnup level that MC calculation and present formula are performed independently. The error between two methods at this level reflects how exactly the Eq.~(\ref{RR6}) fits to the MC reaction rates. The concentration of $^{235}$U changes around $10\%$ from $bu_{1}$ (in Table~\ref{start}) to $bu_{2}$ (in Table~\ref{data2}). The average error $0.05\%$ indicates that the present formula well describes the isotope distribution of $^{235}$U at the burnup level. The average error of the isotope distribution of $^{239}$Pu is around $1\%$. It is partially from the large change of the concentration from $bu_{1}$ to $bu_{2}$ and partially from the coefficient of the $(n,f)$ reaction rate for $^{239}$Pu in Eq.~(\ref{RR6}). The coefficient in Table~\ref{start} is decreasing from $2.50$ to $2.47$. The use of $2.49$ in the Eq.~\ref{RR6} contributes to the $1\%$ error between MC calculation and the present formula at $bu_{2}$.

At the burnup level $bu_{8}$ and $bu_{14}$, the errors become larger, indicating that the reaction rates in Eq.~(\ref{RR6}) change slightly during burnup. The detail will be discussed later. It can be seen that the radial distribution of $N_{9}$ does not change much after a few periods because the production and reaction rates find a balance at such distribution. Both the comparisons of the local burnup and the isotope distributions imply that the present simple and analytical formula can give very nice description if a starting point is given.

\begin{table}
\begin{center}
\caption{\label{data2}  The comparison of the concentrations (unit in cm$^{-3}$) at three burnup levels given by present formula (AN) and MC calculations. }
\vspace{1mm} \setlength{\tabcolsep}{7pt}
\begin{tabular}{cccccccc}
\hline
    $r'/r_{0}$  &  $N_{5,MC}$  &   $N_{8,MC}$ & $N_{9,MC}$ & $N_{5,AN}$ & $N_{8,AN}$ & $N_{9,AN}$   \\
\hline
&$bu_{2}$ &&&&& \\
1/4         & 6.01E+20	& 2.17E+22	& 6.13E+19 	&  6.00E+20	& 2.17E+22	& 6.09E+19    \\
5/8         & 5.98E+20	& 2.17E+22	& 6.46E+19 	&  5.98E+20	& 2.16E+22	& 6.43E+19    \\
13/16       & 5.94E+20	& 2.16E+22	& 7.09E+19 	&  5.94E+20	& 2.16E+22	& 7.08E+19    \\
29/32    	  & 5.91E+20	& 2.16E+22	& 8.13E+19 	&  5.91E+20	& 2.16E+22	& 8.16E+19    \\
61/64    	  & 5.89E+20	& 2.16E+22	& 1.01E+20 	&  5.89E+20	& 2.16E+22	& 1.02E+20    \\
125/128     & 5.88E+20	& 2.16E+22	& 1.31E+20 	&  5.88E+20	& 2.16E+22	& 1.35E+20    \\
127/128     & 5.87E+20	& 2.15E+22	& 1.98E+20 	&  5.87E+20	& 2.15E+22	& 2.02E+20    \\
average errors &&&& 0.05\%	& 0.06\%	& 1.06\% \\
\hline
& $bu_{8}$ &&&&& \\
1/4         & 3.20E+20	& 2.14E+22	& 1.42E+20 	&  3.24E+20	& 2.14E+22	& 1.33E+20    \\
5/8         & 3.14E+20	& 2.14E+22	& 1.49E+20 	&  3.18E+20	& 2.14E+22	& 1.42E+20    \\
13/16       & 3.07E+20	& 2.14E+22	& 1.61E+20 	&  3.10E+20	& 2.14E+22	& 1.53E+20    \\
29/32    	  & 3.01E+20	& 2.13E+22	& 1.84E+20 	&  3.04E+20	& 2.14E+22	& 1.65E+20    \\
61/64    	  & 2.98E+20	& 2.13E+22	& 2.26E+20 	&  3.01E+20	& 2.13E+22	& 2.02E+20    \\
125/128     & 2.96E+20	& 2.11E+22	& 2.93E+20 	&  2.99E+20	& 2.11E+22	& 2.78E+20    \\
127/128     & 2.94E+20	& 2.08E+22	& 4.36E+20 	&  2.97E+20	& 2.09E+22	& 4.06E+20    \\
average errors &&&& 1.07\%	& 0.03\%	& 	7.03\% \\
\hline
& $bu_{14}$ &&&&& \\
1/4         & 1.53E+20	& 2.12E+22	& 1.65E+20 	&  1.60E+20	& 2.12E+22	& 1.52E+20    \\
5/8         & 1.48E+20	& 2.12E+22	& 1.70E+20 	&  1.54E+20	& 2.12E+22	& 1.61E+20    \\
13/16       & 1.42E+20	& 2.11E+22	& 1.83E+20 	&  1.47E+20	& 2.11E+22	& 1.72E+20    \\
29/32    	  & 1.38E+20	& 2.10E+22	& 2.07E+20 	&  1.42E+20	& 2.11E+22	& 1.85E+20    \\
61/64    	  & 1.35E+20	& 2.09E+22	& 2.54E+20 	&  1.39E+20	& 2.09E+22	& 2.23E+20    \\
125/128     & 1.33E+20	& 2.06E+22	& 3.27E+20 	&  1.37E+20	& 2.06E+22	& 3.07E+20    \\
127/128     & 1.32E+20	& 2.01E+22	& 4.79E+20 	&  1.36E+20	& 2.01E+22	& 4.43E+20    \\
average errors &&&& 3.36\%	& 0.07\%	& 	7.89\% \\
\hline
\end{tabular}
\end{center}
\end{table}

\begin{figure}
\includegraphics[scale=0.30]{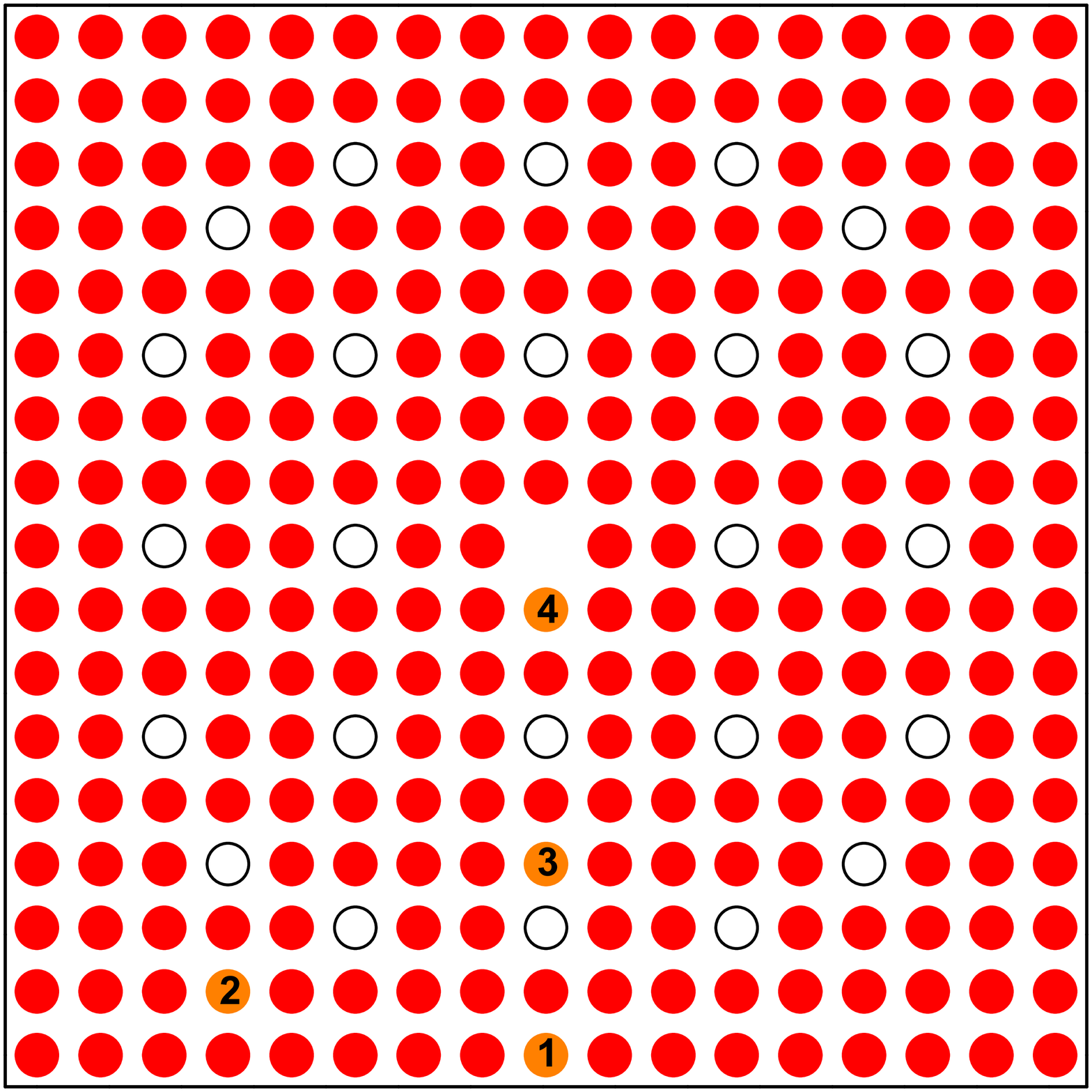}
\caption{\label{ASSM} The model of an assembly, with filled and empty circle for fuel and control rods, respectively. Four selected rods are marked.}
\end{figure}

As the present model is constrained in a single rod, it is worth to examine its validity in a more realistic case, such as in an assembly. Figure~\ref{ASSM} shows the model of an assembly which has the same size for each fuel rod as the previous single one. There are $24$ positions for the control rods and one postion in the center for the detectors. In the present calculation, the control rods are not inserted and the rod for detectors are filled with water. The reaction rates of the fuel rods in the assembly are simulated by setting all fuel rods the same concentrations as the single rod at a certain burnup level. The single rod model are compared with four selected fuel rods in the assmebly, marked in FIG.~\ref{ASSM}.

The relative reaction rates, defined by Eq.~(\ref{RR6}), in the single rod and four rods in the assembly are compared at the burnup level $bu_{2}$ in Table~\ref{data3}. The relative reaction rates of all five rods are generally similar, especially for the two important reactions, $(n,f)$ reaction of $^{235}$U and $^{239}$Pu. The relative reaction rate of $(n,\gamma)$ reaction of $^{238}$U in the single rod is a few percent less than those in the four rods in the assembly. A slight modification can be applied for the Eq.~(\ref{RR6}) to fit to the fuel rods in the assembly. In the case of assembly, it is clear to see that the relative reaction rates are almost constant for all four fuel rods, while the reaction rates are a little different depending on the position of the fuel rods, seen from $\frac{R_{5,f}(rodi)}{R_{5,f}(rod1)}$ in Table~\ref{data3}.

\begin{table}
\begin{center}
\caption{\label{data3} The comparison of the relative reaction rates for a single fuel rod and four selected fuel rods in an assembly at $bu_{2}$. }
\vspace{1mm} \setlength{\tabcolsep}{7pt}
\begin{tabular}{ccccccc}
\hline
    $r'/r_{0}$  &  $R'_{5,f}$& $R'_{8,f}$& $R'_{9,f}$ & $R'_{5,\gamma}$& $R'_{8,\gamma}$ &$\frac{R_{5,f}(rodi)}{R_{5,f}(rod1)}$  \\
\hline
&single&&&&& \\
1/4         &  1.00 & 	0.0026 & 	2.43 & 	0.23 & 	0.0186 &  	    \\
5/8         &  1.02 & 	0.0025 & 	2.43 & 	0.23 & 	0.0192 &  	    \\
13/16       &  1.05 & 	0.0024 & 	2.43 & 	0.23 & 	0.0206 &  	    \\
29/32    	  &  1.07 & 	0.0024 & 	2.42 & 	0.23 & 	0.0235 &  	    \\
61/64      	&  1.08 & 	0.0023 & 	2.42 & 	0.23 & 	0.0289 &  	    \\
125/128     &  1.09 & 	0.0023 & 	2.42 & 	0.23 & 	0.0376 &  	    \\
127/128     &  1.09 & 	0.0023 & 	2.42 & 	0.23 & 	0.0561 &  	    \\
\hline
&rod1&&&&& \\
1/4         &  1.00 & 	0.0025 & 	2.44 & 	0.23 & 	0.0178 & 	    \\
5/8         &  1.02 & 	0.0024 & 	2.43 & 	0.23 & 	0.0185 & 	    \\
13/16       &  1.05 & 	0.0023 & 	2.42 & 	0.23 & 	0.0197 & 	    \\
29/32    	  &  1.07 & 	0.0023 & 	2.42 & 	0.23 & 	0.0222 & 	    \\
61/64      	&  1.08 & 	0.0022 & 	2.42 & 	0.23 & 	0.0269 & 	    \\
125/128     &  1.09 & 	0.0022 & 	2.42 & 	0.23 & 	0.0352 & 	    \\
127/128     &  1.10 & 	0.0022 & 	2.42 & 	0.23 & 	0.0521 & 	    \\
\hline
&rod2&&&&& \\
1/4         &  1.00 & 	0.0025 & 	2.42 & 	0.23 & 	0.0179 & 	0.99     \\
5/8         &  1.02 & 	0.0024 & 	2.42 & 	0.23 & 	0.0184 & 	0.99     \\
13/16       &  1.05 & 	0.0023 & 	2.41 & 	0.23 & 	0.0197 & 	1.00     \\
29/32    	  &  1.07 & 	0.0023 & 	2.40 & 	0.23 & 	0.0225 & 	1.00     \\
61/64      	&  1.09 & 	0.0022 & 	2.40 & 	0.23 & 	0.0277 & 	1.00     \\
125/128     &  1.09 & 	0.0022 & 	2.40 & 	0.23 & 	0.0355 & 	1.00     \\
127/128     &  1.10 & 	0.0022 & 	2.40 & 	0.23 & 	0.0538 & 	1.00     \\
\hline
&rod3&&&&& \\
1/4         &  1.00 & 	0.0023 & 	2.43 & 	0.23 & 	0.0173 & 	1.05     \\
5/8         &  1.02 & 	0.0023 & 	2.43 & 	0.23 & 	0.0178 & 	1.05     \\
13/16       &  1.05 & 	0.0022 & 	2.42 & 	0.22 & 	0.0191 & 	1.06     \\
29/32    	  &  1.07 & 	0.0021 & 	2.41 & 	0.22 & 	0.0220 & 	1.06     \\
61/64      	&  1.09 & 	0.0021 & 	2.41 & 	0.22 & 	0.0271 & 	1.06     \\
125/128     &  1.09 & 	0.0021 & 	2.41 & 	0.22 & 	0.0358 & 	1.06     \\
127/128     &  1.10 & 	0.0021 & 	2.41 & 	0.22 & 	0.0533 & 	1.06     \\
\hline
&rod4&&&&& \\
1/4         &  1.00 & 	0.0023 & 	2.42 & 	0.22 & 	0.0170 & 	1.08     \\
5/8         &  1.03 & 	0.0022 & 	2.42 & 	0.22 & 	0.0175 & 	1.09     \\
13/16       &  1.05 & 	0.0021 & 	2.41 & 	0.22 & 	0.0190 & 	1.09     \\
29/32    	  &  1.07 & 	0.0021 & 	2.40 & 	0.22 & 	0.0216 & 	1.09     \\
61/64      	&  1.08 & 	0.0020 & 	2.40 & 	0.22 & 	0.0268 & 	1.09     \\
125/128     &  1.09 & 	0.0020 & 	2.40 & 	0.22 & 	0.0349 & 	1.09     \\
127/128     &  1.10 & 	0.0020 & 	2.39 & 	0.22 & 	0.0526 & 	1.09     \\
\hline
\end{tabular}
\end{center}
\end{table}

Table~\ref{data4} presents the comparison of the relative reaction rates between the single rod and the rod$1$ in the assembly at the burnup level $bu_{9}$ and $bu_{15}$. As relative reaction rates in the four rods in the assembly are similar to each other, only the data of rod$1$ are shown in Table~\ref{data4}. At both $bu_{9}$ and $bu_{15}$, the $(n,f)$ reaction of $^{235}$U and $^{239}$Pu are similar in both rods, while the relative reaction rate of $(n,\gamma)$ of reaction $^{238}$U in the single rod is a few percents larger than that in the rod$1$. The situations are similar to that at $bu_{2}$. The comparisons in Table~\ref{data3} and \ref{data4} imply that the single rod model is a nice estimation of the fuel rods in the assembly.

\begin{table}
\begin{center}
\caption{\label{data4} The comparison of the relative reaction rates for a single fuel rod and the fuel rod in an assembly at $bu_{9}$ and $bu_{15}$. }
\vspace{1mm} \setlength{\tabcolsep}{7pt}
\begin{tabular}{cccccc}
\hline
    $r'/r_{0}$  &  $R'_{5,f}$& $R'_{8,f}$& $R'_{9,f}$ & $R'_{5,\gamma}$& $R'_{8,\gamma}$   \\
\hline
&$bu_{9}$ single&&&& \\
1/4         &  1.00 & 	0.0025 & 	2.23 & 	0.23 & 	0.0183  	  \\
5/8         &  1.02 & 	0.0025 & 	2.24 & 	0.23 & 	0.0190  	  \\
13/16       &  1.04 & 	0.0024 & 	2.25 & 	0.23 & 	0.0205  	  \\
29/32    	  &  1.06 & 	0.0023 & 	2.26 & 	0.23 & 	0.0233  	  \\
61/64      	&  1.07 & 	0.0023 & 	2.26 & 	0.23 & 	0.0287  	  \\
125/128     &  1.08 & 	0.0023 & 	2.27 & 	0.23 & 	0.0374  	  \\
127/128     &  1.08 & 	0.0023 & 	2.27 & 	0.23 & 	0.0559  	  \\
\hline
&$bu_{9}$ rod1&&&& \\
1/4         &  1.00 & 	0.0024 & 	2.23 & 	0.23 & 	0.0175      \\
5/8         &  1.02 & 	0.0024 & 	2.24 & 	0.23 & 	0.0182      \\
13/16       &  1.04 & 	0.0023 & 	2.25 & 	0.23 & 	0.0197      \\
29/32    	  &  1.06 & 	0.0022 & 	2.26 & 	0.23 & 	0.0222      \\
61/64      	&  1.07 & 	0.0022 & 	2.26 & 	0.23 & 	0.0272      \\
125/128     &  1.07 & 	0.0022 & 	2.27 & 	0.23 & 	0.0354      \\
127/128     &  1.08 & 	0.0022 & 	2.28 & 	0.23 & 	0.0523      \\
\hline
&$bu_{15}$ single&&&& \\
1/4         &  1.00 & 	0.0022 & 	2.18 & 	0.23 & 	0.0168      \\
5/8         &  1.02 & 	0.0022 & 	2.19 & 	0.23 & 	0.0174      \\
13/16       &  1.04 & 	0.0021 & 	2.20 & 	0.22 & 	0.0188      \\
29/32    	  &  1.05 & 	0.0021 & 	2.21 & 	0.22 & 	0.0214      \\
61/64      	&  1.06 & 	0.0021 & 	2.22 & 	0.22 & 	0.0265      \\
125/128     &  1.06 & 	0.0020 & 	2.22 & 	0.23 & 	0.0346      \\
127/128     &  1.07 & 	0.0020 & 	2.23 & 	0.23 & 	0.0510      \\
\hline
&$bu_{15}$ rod1&&&& \\
1/4         &  1.00 & 	0.0021 & 	2.18 & 	0.22 & 	0.0161      \\
5/8         &  1.02 & 	0.0021 & 	2.19 & 	0.22 & 	0.0166      \\
13/16       &  1.04 & 	0.0020 & 	2.20 & 	0.22 & 	0.0178      \\
29/32    	  &  1.05 & 	0.0020 & 	2.20 & 	0.22 & 	0.0203      \\
61/64      	&  1.06 & 	0.0020 & 	2.22 & 	0.22 & 	0.0251      \\
125/128     &  1.06 & 	0.0020 & 	2.21 & 	0.22 & 	0.0327      \\
127/128     &  1.07 & 	0.0019 & 	2.22 & 	0.22 & 	0.0486      \\
\hline
\end{tabular}
\end{center}
\end{table}

It is worth to consider why such simple formula Eq.~(\ref{RR6}) works. Take $R'_{9,f}(r)$ for example,
\begin{eqnarray}\label{RR10}
R'_{9,f}(r) &=& \frac{R_{9,f}(r)/N_{9}(r)}{R_{5,f}(r)/N_{5}(r)} \nonumber \\
 &=& \frac{\int_{E} \phi(r,E)\sigma(9,E)dE}{\int_{E} \phi(r,E)\sigma(5,E)dE}, \nonumber
\end{eqnarray}
which is the ratio of the one-group effective cross section. An approximately constant $R'_{9,f}(r)$ indicate a constant neutron dynamics for the ratio between the fission reaction of $^{239}$Pu and $^{235}$U along the radial direction. Although the one-group effective cross section changes as the function of $r$, the ratio of the one-group effective cross section of each reaction included in the present study keeps almost the same except the $(n,\gamma)$ reaction of $^{238}$U. One of the advantage of the present method is the consideration of the reaction rate compared with the previous methods~\cite{Schubert2008,Lee20002,Soba2013}. Because the local burnup and the isotope distributions do not directly depend on the neutron flux but on the reaction rate, it is an alternative method which consider the properties of the latter one. The TRANSURANUS model considers the constant neutron flux and the one-group effective cross sections during $\Delta bu$. The assumption in the present model is actually the constant ratio of the one-group effective cross sections, which is relatively easy to be achieved.

From the Table~\ref{start}, \ref{data3}, and \ref{data4}, it is seen that the relative reaction rates changes not much along the radial direction and among each burnup level. It should be mentioned that the relative $(n,\gamma)$ reaction rates of $^{239}$Pu are also close to constant. It is neglected in the present calculation because it links to other Pu isotopes which are not included in the present study. If the error of a few percents is acceptable, the Eq.~(\ref{RR6}) can be performed independently until $bu_{15}$ because the relative reaction rates are almost independent both on radius (except the $(n,\gamma)$ reaction of $^{238}$U) and burnup. More exact investigations need consider that the relative reaction rates are not exactly constant as the burnup increasing. Such as from $bu_{2}$ to $bu_{9}$, $R'_{9,f}$ decreases from $2.49$ to $2.25$, seen from Table~\ref{data3} and \ref{data4}. This is why the difference of isotope distributions between the MC and present calculations becomes larger as the increment of the burnup. One can use the present formula to obtain the concentration at a certain burnup level, such as $bu_{4}$, re-simulate the reactions with the concentrations through MC, and refit the coefficients in the Eq.~(\ref{RR6}).

The present calculation is limited to $^{235}$U, $^{238}$U, and $^{239}$Pu to examine its validity in a simple case. It is expected that it can be expanded to more complicated case, such as the consideration of other Pu isotopes and other nuclides which played important role in the reactor, such as the minor actinide and the poisons. In that case, a starting point is also needed to obtain the parameters in the Eq.~(\ref{RR6}). In addition, the present calculation constrains the $\Delta bu$ to compare with the MC calculations. In principle, the $\Delta bu$ can be rather small to obtain an almost continues changes of the reaction rates depending on the burnup and radius. In the case of small $\Delta bu$, the assumption of constant reaction rates in Eq.~(\ref{deltabuRR}) is more reasonable.

For a more exact study, the present method can be used together with the MC calculation. Such as, starting from $bu_{1}$, the reaction rates and the concentrations can be calculated with rather small $\Delta bu$ up to $bu_{2}$. The output is used for the next MC calculation. If the $\Delta bu$ is rather small, purely MC calculation is very time consuming from $bu_{1}$ to $bu_{2}$. But with the supplement of the present work, more exact and relatively quick results are expected.

In practice, one can use the reaction rates in Eq.~(\ref{RR6}) with the concentrations at $bu_{1}$ to calculate the $\Delta N(r)$ in Eq.~(\ref{deltabuRR}) with small $\Delta bu$, such as $0.01$ MWd/kgU. The next step is to use the new concentrations to calculate the reaction rates in Eq.~(\ref{RR6}) and then the next $\Delta N(r)$ through Eq.~(\ref{deltabuRR}). After $285$ steps, the concentrations at $bu_{2}$ are obtained and can be used for the next MC calculation. It is actually the Euler method with very small steps, which is expected to be more exact. In Eq.~(\ref{deltabuRR}), the reaction rates (actually the neutron flux and the concentrations) are assumed to be constant during $\Delta bu$. But both the neutron flux and the concentrations change when the burnup changes. For detail, the concentration is in the reaction rate in the numerator of the Eq.~(\ref{deltabuRR}). The effect of the neutron flux in Eq.~(\ref{deltabuRR}) is reflected by the integral of the reaction rates (the total fission reaction rate) in the denominator. At the different burnup level, the total fission rate and its distribution in each isotopes are different, which corresponds to the different neutron flux. If the calculation between $bu_{1}$ and $bu_{2}$ are separated to many small steps, the neutron flux and concentrations are reconsidered for each steps.

The present method is also helpful for solving the properties of the fuel rods at the different positions in an assembly or an reactor core. As discussed before, the relative reaction rates are almost constant in the fuel rod at the different position in the assembly. Only one factor is needed to describe the reaction rates in each fuel rod. One can simulate at an initial level through the MC method to obtain the factors for each fuel rod and investigate the evolution of the local burnup and isotope distributions for all fuel rods at certain average burnup level.

\section{\label{sec:level4}Summary}
In conclusion, an analytical and simple formula is suggested to calculate the reaction rates as well as the burnup and the isotope distributions as the function of radius. The parameters of the formula are fitted to the reaction rates of a given burnup level. Starting from the same burnup level, the present formula can give very nice description compared with a MC calculation from the code TRIPOLI-4. The reaction rates depend on the concentrations and the neutron flux, which both vary along the radial direction and at the different burnup levels. The present work finds almost constant relative reaction rates on these two degrees of freedom (except the $(n,\gamma)$ reaction of $^{238}$U on the degree of freedom of the radius), which provides a solid physical explanation on the simple formula used to calculate the local burnup and the isotope distributions.

\section*{Conflicts of Interest}
The authors declare that there is no conflict of interest regarding the publication of this manuscript.

\section*{Acknowledgement}
The authors acknowledge to our collaborator CEA for the authorization of the TRIPOLI-4, and the useful discussions with Sicong Xiao, Min Zhang, Youjun Fang, and Nianwu Lan. This work has been supported by the National Natural Science Foundation
of China under Grant No.~11305272, the Specialized Research Fund for the Doctoral Program
of Higher Education under Grant No.~20130171120014, the Guangdong Natural Science Foundation under Grant No.~2014A030313217, and the Pearl River S\&T Nova Program of Guangzhou under Grant No.~201506010060.

\bibliographystyle{elsarticle-num}

\end{document}